\documentclass[twocolumn,pre]{revtex4}
\usepackage{graphicx}
\usepackage{graphicx}
\usepackage{dcolumn}
\usepackage{bm}
\usepackage{color}
\usepackage{amsmath}
\usepackage{amsfonts}
\usepackage{amssymb}

\begin{document}
\title{Effects of static and temporally fluctuating tensions on semiflexible polymer looping}

\author{Jaeoh Shin and Wokyung Sung}
\affiliation{Department of Physics and PCTP, Pohang University of Science and Technology, Pohang 790-784, South Korea}
\date{October 24, 2011}

\begin{abstract}
Biopolymer looping is a dynamic process that occurs ubiquitously in cells for gene regulation, protein folding, etc. In cellular environments, biopolymers are often subject to tensions which are either static, or temporally fluctuating far away from equilibrium. We study the dynamics of semiflexible polymer looping in the presence of such tensions by using Brownian dynamics simulation combined with an analytical theory. We show a minute tension dramatically changes the looping time, especially for long chains. Considering a dichotomically flipping noise as a simple example of the nonequilibrium tension, we find the phenomenon of resonant activation, where the looping time can be the minimum at an optimal flipping time. We discuss our results in connection with recent experiments.
\end{abstract}

\maketitle
\section{Introduction}
 The kinetics of polymer loop formation has been studied for several decades and recently has attracted renewed attention due to the particular importance in biology. The DNA loop formation is a basic process that underlies genetic expression, replication, and recombination~\cite{schleif1992,matthews1992}. For example, in \textit{E. coli} the \textit{lac} repressor (LacI)-mediated loop is crucial for the repressive regulation of \textit{lac} genes. The hairpin loop formation is the elementary step in protein folding~\cite{thirumalai1999} and structure formation in RNA folding~\cite{tinoco1999}.  
 
 A cell is crowded with a multitude of subcellular structures including globular proteins and RNAs~\cite{minton2001}, with which DNA is constantly interacting. A DNA fragment about to loop is often subject to temporally fluctuating forces due to its dynamic environment including the other part of the chain. Recently, the power spectrum of the fluctuating force exerted on cytoskeleton was measured to be an order of magnitude larger than that expected from thermal equilibrium condition~\cite{gallet2009}. This indicates that the cell interior is an active and nonequilibrium medium.
 
 The advance of single molecule experiment techniques provides detailed information on the DNA loop formation. Finzi and Gelles~\cite{finzi1995} observed LacI-mediated DNA loop formation and dissociation by monitoring nano-scale Brownian motion of the micron-sized particle attached to one end of the DNA. Lia \textit{et al.}~\cite{lia2003} showed that in \textit{gal} repressor and DNA-bending protein HU mediated looping, mechanical constraints such as tension and torsion play a pivotal role. Gemmen~\textit{et al.}~\cite{gemmen2006} studied effects of tension in the presence of two-site restriction enzymes which can cut the DNA upon binding on two sites simultaneously. They found that the cleavage activity decreases approximately 10-fold as the tension increases from 30 fN to 700 fN. They also found that the optimum loop size decreases with the tension, which is qualitatively in agreement with theoretical predictions~\cite{sankararaman2005}. More recently, Chen \textit{et al.}~\cite{chen2010a} studied effects of tension in femtonewton range on the kinetics of LacI-mediated DNA looping. They found that small tension of 100 fN scale on the substrate DNA can not only increases the looping time~\cite{chen2010a} but also found that the looping time is greatly reduced in the presence of a fluctuating tension~\cite{chen2010b}. These results suggest the ubiquitous roles of the static and temporally fluctuating tensions in regulation of the DNA loop formation. Yet, there appears to be no unifying conceptual or theoretical framework that explains a variety of experiments including these.

Theoretically, on the other hand, Yan \textit{et al.}~\cite{yan2005} developed a transfer matrix method to calculate semiflexible polymer end-to-end distance distribution function and loop formation probability (or $J$-factor). They studied various effects of nonlinear elasticity arising from DNA bending protein-induced kinks~\cite{lia2003} or thermal-fluctuation-induced bubbles on the $J$-factor. Their study provides a valuable insight to understand DNA bending on short length scale~\cite{yan2005}, which has attracted much attention recently \cite{cloutier2004,du2005}. They also studied effects of tension on the $J$-factor~\cite{yan2005}, which is related to the free energy barrier for loop formation~\cite{jun2003}, thus to the loop formation rate. Similar results are obtained by using an elastic theory of a semiflexible polymer~\cite{sankararaman2005}. However, since the loop formation rate is not proportional to the $J$-factor alone but depends on the free energy given an arbitrary chain end-to-end distance, it is hard to quantitatively compare these theories to the experiment~\cite{chen2010a}.

Independently, Blumberg \textit{et al.}~\cite{blumberg2005} studied effects of static tension on protein-mediated DNA loop formation by modeling the DNA conformation to be either one of two states, looped and unlooped states. In their appealing calculations of free energy change associated with the transition under the tension, they considered not only the stretching free energy of DNA but also DNA alignment constraint imposed by protein binding. They found that for the loop size larger than 100 base pair distance ($\approx34$ nm), a tension of 0.5 pN can increase the looping time by more than two order of magnitude. 
There is room for improvement in their approach, however, on the evaluation of the free energy that can be valid for short end-to-end distance of the chain as well as a description of detailed kinetic process using the mean first-passage time approach.

In this paper, in an effort to understand the basic physical mechanism of the biopolymer looping in a coherent manner, we perform Brownian dynamics simulation of semiflexible polymers treated as extensible wormlike chain, combined with one-dimensional theory of barrier crossing over the free energy of loop formation. For analytical understanding, we use, as an example, the mean-field wormlike chain model~\cite{thirumalai1998}, which is shown to be a good approximation for the free energy for the chain lengths we consider here. With static tensions, we find that the looping time, defined as the mean first-passage time to cross the free energy barrier, steeply increases with the applied tension $f$, in an agreement with our simulation results but distinct from the previous theoretical result~\cite{blumberg2005}. For the case of time-dependent tension, we consider dichotomically fluctuating tension, where the looping times are found to be reduced, consistent with the experiment~\cite{chen2010b}. Most importantly, we find so-called the resonant activation, where the looping time is the minimum at an optimal flipping time of the dichotomic force. In this exploratory study, we neglect the alignment constraint on the loop formation, which is minor effect for the chain lengths we consider here~\cite{blumberg2005}. 

In the following section, we describe our polymer model and simulation method, whose results are discussed in Sec.~\ref{sec:results}. We summarize our results in Sec.~\ref{sec:conclusion}.

\section{Polymer Model and Simulation Method}
\label{sec:model}
We consider the semiflexible polymer looping in the presence of static and fluctuating tension. The polymers are modeled as semiflexible chains of $N$ beads of diameter $d$, with the interaction potential $U$. Here $U=U_{s}+U_{b}$, where $U_{s}$ and $U_{b}$ are the stretching and bending energy
\begin{eqnarray}
U_{s}= \frac{k}{2}\sum_{i=1}^{N-1} (|\vec{r}_{i+1}-\vec{r}_{i}|-l_0)^2,
\end{eqnarray}
\begin{eqnarray}
U_{b}= \frac{\kappa}{2}\sum_{i=2}^{N-1} \theta_{i}^2,
\end{eqnarray}
 where $k$ is the stretching stiffness, $l_0$ is the natural bond length, $\kappa$ is the bending stiffness, $\vec{r}_{i}$ is the position of the $i$th bead, and $\theta_i$ is the angle of the $i$th bond. 
\begin{figure}
\vspace{0.3cm}
\includegraphics[width=8.5cm]{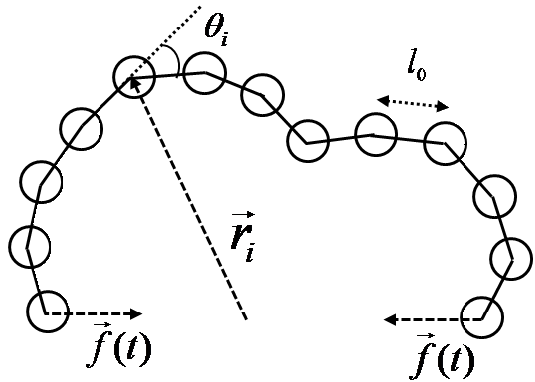}
\caption{A schematic picture of the model semiflexible polymer under a negative tension $f(t)$.\label{fig:schematic}}
\end{figure}
The dynamics of $N$ beads ($i=1, 2, ..., N$) are described by the overdamped Langevin equation
\begin{eqnarray}
\gamma \dot{\vec{r}}_{i}(t)=-\nabla_{i}U + \vec{\xi}_{i}(t)+\vec{f}(t)(\delta_{i,N}-\delta_{i,1})\label{eq:eqmotion},
\end{eqnarray}
where $\gamma$ is the friction coefficient of the bead, and $\vec{\xi}_{i}$ is the Gaussian random noise with mean $\langle\vec{\xi}_{i}(t)\rangle=0$ and variance $\langle\xi_{i,p}(t)\xi_{j,q}(0)\rangle=2\gamma k_{B}T\delta_{i,j}\delta_{p,q}\delta (t)$. Here, $\langle...\rangle$ denotes ensemble average, $p$ and $q$ are the Cartesian coordinate indices, $k_B$ is the Boltzmann constant, and $T$ is the absolute temperature.
The additional force $\vec{f}(t)$ is the tension which is applied only to the two end segments ($i=1$ and $N$) along the chain end-to-end direction, $\vec{f}(t)=f(t) \dfrac{\vec{r}_N-\vec{r}_1}{|\vec{r}_N-\vec{r}_1|} $ (see Fig.~\ref{fig:schematic}). The bending stiffness of the short semiflexible chains we consider allows us to neglect the excluded volume effect. Furthermore, we neglect the hydrodynamic interactions, which is found to be small in the short chains we consider here~\cite{explain}.

We use the parameters $l_0$, $k_{B}T$, and $\gamma l_{0}^2/(k_{B}T)$ to fix the length, energy, and time scales, respectively. The dimensionless parameters in our simulation are $l_{0}=1$, $k_{B}T=1$, $k=100$, and $\kappa=5$. In our model, we consider $l_{0}$ as 10 nm, so with $\kappa=5$ the persistence length $l_p=\kappa l_{0}/(k_{B}T)$ of the chain is 50 nm. The diameter of the bead is set to be 5 nm to fit the hydrodynamic friction coefficient of a cylinder with 2 nm diameter and 10 nm length, which is about $4.61 \times 10^{-11}$ $\rm{kg}$ $\rm{s^{-1}}$ with the water viscosity $\eta=0.001$ $\rm{kg}$ $\rm{m^{-1}}$ $\rm{s^{-1}}$ at room temperature. The time unit is then 1.14~$\mu$s. 
 The equations of motion are integrated by using a second-order stochastic Runge-Kutta algorithm~\cite{honeycutt1992} with the time-step $\Delta t=2\times10^{-3}$. In our simulation, we consider that the looping occurs whenever the chain end-to-end distance $r=|\vec{r}_{N}-\vec{r}_{1}|$ is shorter than a cutoff distance $l_{c}$. The average were taken over at least 2,000 and 5,000 independent runs for static and fluctuating tension cases, respectively.

\section{Results and Discussion}
\label{sec:results}
\subsection{Effects of static tension on the looping}

\begin{figure}
\vspace{0.3cm}
\includegraphics[width=8.5cm]{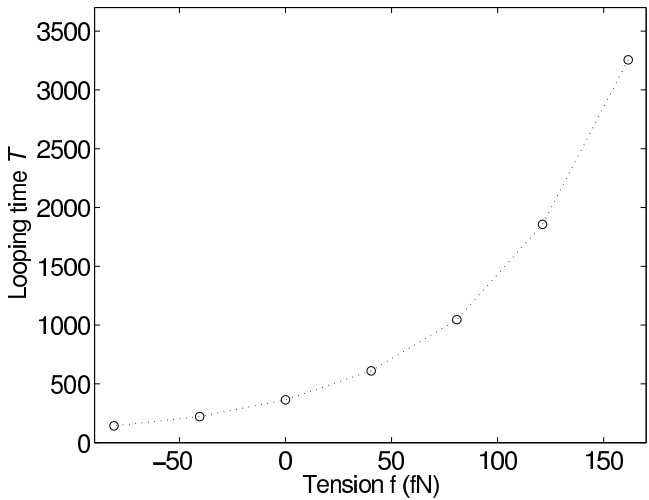}
\caption{The looping time $\mathcal{T}(f)$ as a function of tension $f$ for the chain of $N=12$. \label{fig:N12_Static1}}
\end{figure}

We first study the effects of static tension on semiflexible polymer looping. Here we consider the bead number $N=12$, 18, and 24 which respectively correspond to the chain lengths $L=110$, 170, and 230 nm, similar to the loop size considered in recent experiments~\cite{chen2010a, chen2010b}. Figure~\ref{fig:N12_Static1} shows the looping time $\mathcal {T}$ as a function of tension $f$ for the chain of $N=12$ and $l_{c}=2 l_{0}$. We apply the tension $f$ from -81 to 162 fN, comparable in magnitude to the typical entropic force on the double-stranded DNA, $k_{B}T/l_{p}$ $\approx$ 80 fN. The $f$ is much smaller than the piconewton scale forces those are typically involved in active molecules~\cite{blumberg2005jbp} e.g., molecular motors in a cell. As $f$ changes over the scale as small as 100 fN, the looping time $\mathcal{T}$ dramatically changes; for example, $\mathcal{T}$ increases about 5 times as the tension $f$ increases 120 fN. 

\begin{figure}
\vspace{0.3cm}
\includegraphics[width=8.5cm] {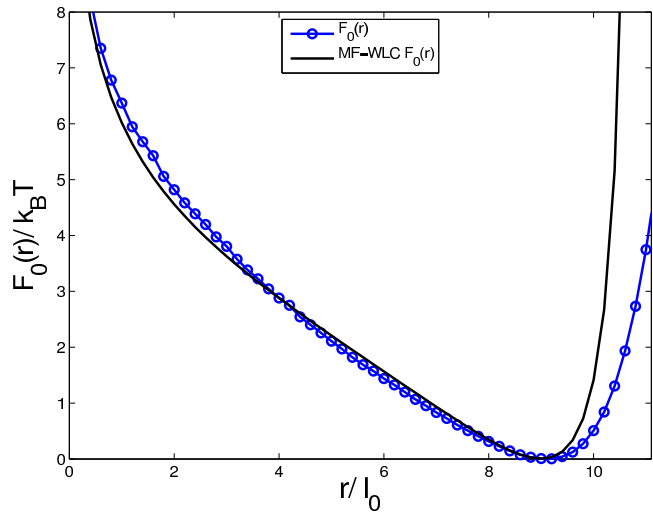}
\caption{The free energy $F_{0}(r)$ of $N=12$ chain obtained from the simulation (solid line with circles) and from the mean-field wormlike chain model~\cite{thirumalai1998} (solid line).\label{fig:N12_mfwlc}}
\end{figure}

To understand this behavior, we consider semiflexible polymer looping as a one-dimensional barrier crossing process~\cite{jun2003}, which is described by the Langevin equation

\begin{eqnarray}
\frac{\gamma} {2} \dot{r}(t)=- \frac{\partial F(r)}{\partial r} + \xi(t)\label{eq:1d_eqmotion}.
\end{eqnarray}
Here, $F(r)=-k_{B}T\log P(r)$ is the free energy of the chain given the chain end-to-end distance $r$, where $P(r)$ is the radial distribution function, and $\xi(t)$ is a random force due to thermal fluctuation given by a Gaussian and white noise that satisfies $\langle\xi(t)\rangle=0$ and $\langle\xi(t)\xi(0)\rangle=\gamma k_{B}T\delta (t)$. The shape of $F(r)$ obtained by the $N$-beads simulation is shown in Fig.~\ref{fig:N12_mfwlc} (solid line with circles) for the chain of $N=12$. In this one-dimensional description, the looping time is the mean first-passage time (MFPT) for the variable $r$ to reach the cutoff distance $l_{c}$ starting from the initial chain end-to-end distance $r_{0}$. It is given by~\cite{risken}

\begin{eqnarray}
\mathcal{T}(r_{0})=\int_{l_c}^{r_0} dr\exp(\beta F(r))\frac{1}{D}\int_{r}^{L} dr'\exp(-\beta F(r')) \label{eq:mfpt},
\end{eqnarray}
where $D=2k_{B}T/ \gamma $ is the relative diffusion coefficient of two end beads, $\beta=(k_{B}T)^{-1}$. Then the looping time is $\mathcal{T}=\langle\mathcal{T}(r_{0})\rangle_{\rm{eq}}=\int_{}^{} dr_{0} \exp(-\beta F(r_{0}))\mathcal{T}(r_0)/\int dr_{0} \exp(-\beta F(r_{0}))$, where $\langle...\rangle_{\rm{eq}}$ represents the average over the initial equilibrium  distribution. The MFPT in the presence of tension $f$, also can be calculated by using the free energy, $F(r)=F_{0}(r)-fr$, where $F_{0}(r)$ is the free energy without tension.

For the $P(r)$, we use the mean-field wormlike chain model (MF-WLC)~\cite{thirumalai1998} as an example. The radial distribution function $P(r)$ of this model is $P(r)\sim r^{2}[1-(\frac{r}{L})^2]^{-\frac{9}{2}} \exp[-\frac{3L}{4 l_{p}} \frac{1}{(1-(r/L)^2)}]$. This formula yields a reasonable approximation to our simulation result for $F_{0}(r)$, except for $r\approx L$, as shown in Fig.~\ref{fig:N12_mfwlc}. The large deviation between two curves in the region near $r/L=1$ is because our model allows the chain extensibility while MF-WLC does not. 
However, according to the single molecule experiment using optical tweezers~\cite{bustamante1996}, the DNA is extensible, manifesting overstretching transition subject to a strong tension, thus showing a significant deviation from the inextensible WLC. The optimal value $k$ to fit the force-extension curve is found to be $k\approx$ 2700~\cite{bustamante1996}. However, the simulation using $k$ as large as $k=2700$ demands computing time much longer than that using $k=100$. Unlike the free energy, the normalized looping time $\mathcal{T}_N(f)/\mathcal{T}_N(0)$, the looping time in the presence of tension $f$ relative to the looping time in its absence, is expected to be quite insensitive to the value of $k$~\cite{vologodskii2000,jian1997}. For this reason we adopt $k=100$, which was also employed in a number of studies on the looping~\cite{vologodskii2000,hyeon2006}.

\begin{figure}
\vspace{0.3cm}
\includegraphics[width=9.5cm]{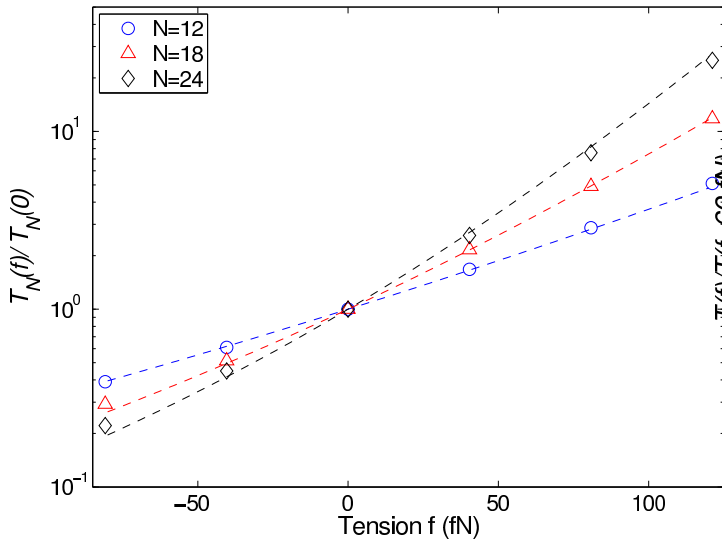}
\caption{The normalized looping time $\mathcal{T}_{N}(f)/\mathcal{T}_{N}(0)$ as a function of tension $f$ for $N=12$, $18$ and $24$. The symbols are simulation data and the dashed lines are from mean first-passage time calculations using mean-field wormlike chain model~\cite{thirumalai1998}.\label{fig:Normalized_looping_static}}
\end{figure}

To understand the effect of tension on the looping time, the MF-WLC provides a useful analytical model. Figure~\ref{fig:Normalized_looping_static} shows $\mathcal{T}_N(f)/\mathcal{T}_{N}(0)$ as a function of $f$ for the chain lengths $N$=12, 18, and 24. Here the symbols are simulation results and the dashed lines are from MFPT calculations (Eq.~\ref{eq:mfpt}). For the range of $f$ we study here, two results are in an excellent agreement. This figure also shows that the normalized looping time increases exponentially with $f$. The reason is that, for the short chain lengths considered here, the free energy of the loop formation increases with $f \langle r\rangle_{\rm{eq}} \approx fL$, so $\mathcal{T}$ increases approximately in exponential with $f$. In contrast, the theory of Ref.~\cite{blumberg2005} considered that, for low tension ($f<80$ fN), $\mathcal{T}$ increases exponentially with $f^{2}$ because of Gaussian force-extension relation they used.

The looping time in the presence of a tension $f$ can be written in dimensionless form
\begin{eqnarray}
\frac{D\mathcal{T}(f)}{L^2}=\bigg\langle \int_{\frac{l_c}{L}}^{\frac{r_0}{L}} dx\int_{x}^{1} dx'e^{\beta (F_{0}(x)-F_{0}(x'))}e^{\beta f L(x'-x)} \bigg\rangle_{\rm{eq}}=\varphi(\beta fL, \frac{l_{c}}{L}, \frac{l_{p}}{L}) 
 \label{eq:dimensionless_mfpt},
\end{eqnarray}
where $x\equiv r/L$. $\varphi$ is a function of a dimensionless scaling variable $\beta f L$. It means for very large $L$, a minute tension $f$ can dramatically change the looping time. Indeed, Fig.~\ref{fig:Normalized_looping_static} shows that, for longer chains, the normalized looping time changes more sensitively with $f$. If $N$ is 100, corresponding to $L\approx1$ $\mu$m, a change of the force as small as 4 fN can affect the looping time. This is a consequence of the cooperativity of the long polymer chains arising from the chain connectivity, which is previously addressed in a study of polymer translocation through membrane~\cite{sung1996}.
The sensitivity of looping on tension is an emergent behavior that manifests beyond the complexity of real DNA loop formation~\cite{allemand2006}, e.g., the details of chain conformations outside of the loop, orientation constraint, and associated proteins.

\begin{figure}
\vspace{0.3cm}
\includegraphics[width=8.5cm]{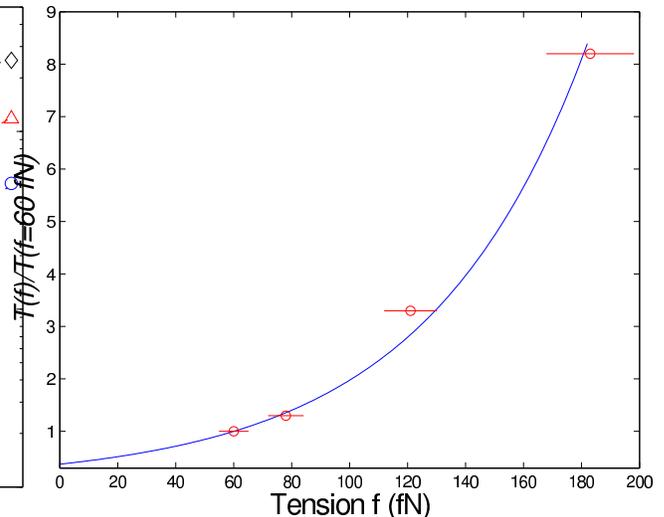}
\caption{The relative looping time $\mathcal{T}(f)/\mathcal{T}$($f=60$ fN) as a function of tension $f$ with the cutoff distance $l_c=5$ nm for $L=103.7$ nm obtained by the MFPT calculation using MF-WLC model~\cite{thirumalai1998} (solid line). The circles are experimental data including error bar from Ref.~\cite{chen2010a}.\label{fig:cutoff_5}}
\end{figure}

Finally, we compare our result with the recent experimental data~\cite{chen2010a} for the DNA loop size $L=103.7$ nm. As shown by the circles in Fig.~\ref{fig:cutoff_5}, the looping time rises steeply as a function of tension $f$ relative to the one with $f=60$ fN, which is the minimum tension used in the experiment. We also plot the corresponding relative MFPT by a solid curve. With the cutoff distance chosen to be the bead diameter, $l_{c}$=5 nm, our result is in good agreement with the experiment~\cite{chen2010a}. Because our model does not consider many details of DNA loop formation, this agreement could be somehow fortuitous but encouraging. A theory including effectively the complexity of DNA loop formation and further controlled experiments of various loop sizes are needed for quantitative comparison.

\subsection{Effects of dichotomically fluctuating tension on the looping}

Now suppose that tension temporally fluctuates due to the nonequilibrium noise inherent \textit{in vivo} systems which can be generated by a variety of constituents of a cell, e.g., protein like RNA polymerase. Also a DNA fragment about to loop is influenced by the other part of the chain whose conformation is constantly fluctuating. As a simple example of nonequilibrium fluctuations, we consider a dichotomic noise, with which the tension $f(t)$ flips between two level of the forces $+f_d$ and $-f_d$ with a mean flipping time $\tau$. The $f(t)$ is a noise with zero mean and its time correlation function is $\langle f(t)f(t')\rangle=f_{d}^{2}\exp(-2|t-t'|/\tau)$. We generate dichotomic noise $f(t)$ using the algorithm described in~\cite{barak2006}. In the initial equilibration time, $f(t)$ is given with either $+f_d$ or $-f_d$ with an equal probability 1/2, and in a small time-step $\Delta t$, $f(t)$ can flip to the other value with the probability $\frac{1}{2}-\frac{1}{2}\exp(-2\Delta t/ \tau)$, which makes the mean flipping time be $\tau$.
\begin{figure}
\includegraphics[width=8.5cm] {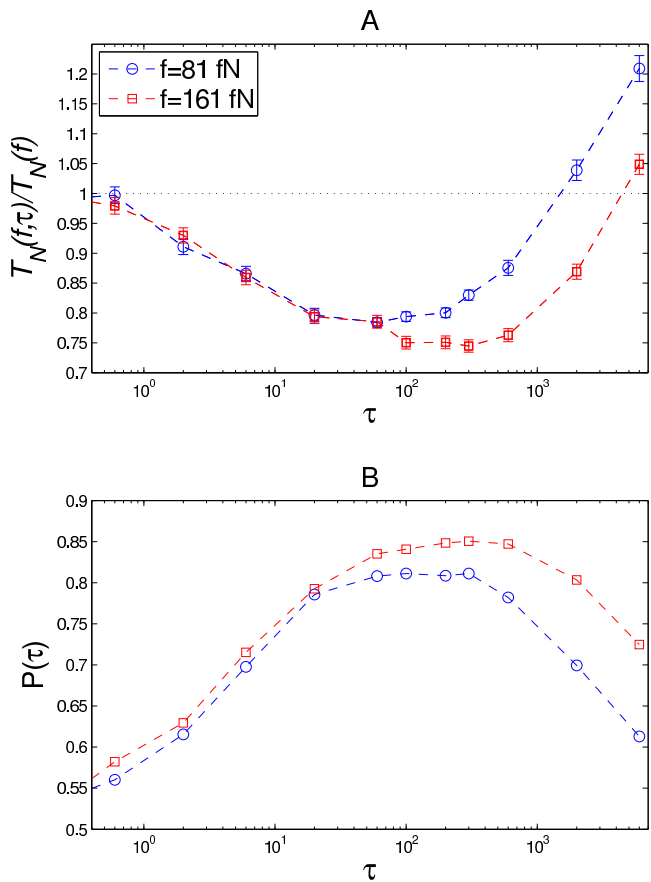}
\caption{(A) The normalized looping time $\mathcal{T}_{N} (f,\tau)/\mathcal{T}_{N} (f)$ in the presence of dichotomic tension as well as static tensions for the chain of $N=12$. The dichotomic force amplitude is $f_d=61$ fN, and the static tensions amplitudes are $f=81$ (circle) and $f=161$ fN (rectangular). The looping time shows a minimum at optimal flipping time $\tau_{r}\approx60$ and $\approx300$ for $f=81$ and 161 fN, respectively. (B) The probability $P(\tau)$ that the dichotomic force is negative (i.e., inward direction) at the instant of looping as a function of $\tau$. \label{fig:RAN12_static}}
\end{figure}

Figure~\ref{fig:RAN12_static} $A$ shows the normalized looping times, $\mathcal{T}_{N} (f,\tau)/\mathcal{T}_{N} (f)$, as a functions of $\tau$ for the chain length $N$=12 in the presence of fluctuating tension $f(t)$ added to the static tension $f$. We consider a value of dichotomic noise amplitude, $f_d=61$ fN and two values of static tensions amplitudes, $f=81$ (circles) and 161 fN (squares), which are similar to those used in a recent experiment~\cite{chen2010b}. For very short $\tau$, the dichotomic forces are averaged out, so the looping times converge to the values without dichotomic force. They gradually decrease with $\tau$ until the minimum at $\tau_r\approx60$ and $\approx300$ for $f=81$ and 161 fN, respectively. For very long $\tau$, the dichotomic force rarely changes in a typical looping time, so the looping time goes to the average of the looping time with tension ($f+f_d$) and the looping time with tension ($f-f_d$). The average is dominated by the looping time with ($f+f_d$), so that the looping time sharply rises with $\tau$.

Related to this, we study the probability $P(\tau)$ that the dichotomic tension $f(t)$ is negative (i.e., is in inward direction) at the instant of looping as a function of $\tau$ (Fig.~\ref{fig:RAN12_static} $B$). It has a maximum at $\tau\approx100$ and $\tau\approx$ 300 for $f=81$ and 161 fN, respectively, which means that, for $\tau$ near the $\tau_{r}$, most of looping occurs when $f(t)$ is in inward direction, and the looping time becomes the minimum. In the limits of very short or very long $\tau$, $f(t)$ is positive or negative with equal probabilities at the moment of looping. The maximum of $P(\tau)$ is larger for larger $f$. Evidently the minimum looping time is closely associated with the maximum of the $P(\tau)$.

The minimum of $\mathcal{T}$ is found to occur when the flipping time $\tau$ is comparable to the diffusion time of a Brownian \textit{particle} in the free energy $F(r)$, $\tau_{r} =\alpha L^{2}/D$, where $\alpha$ is a numerical value of order unity that increases with the tension $f$~\cite{doering1992}. This phenomenon is an extension of the resonant activation (RA) originally found in the single Brownian particle crossing over a fluctuating barrier~\cite{doering1992}. Indeed, we can regard the polymer looping in the presence of dichotomic tension as the process of a Brownian \textit{particle} crossing over a fluctuating \textit{free energy} barrier.

\begin{figure}
\includegraphics[width=8.5cm] {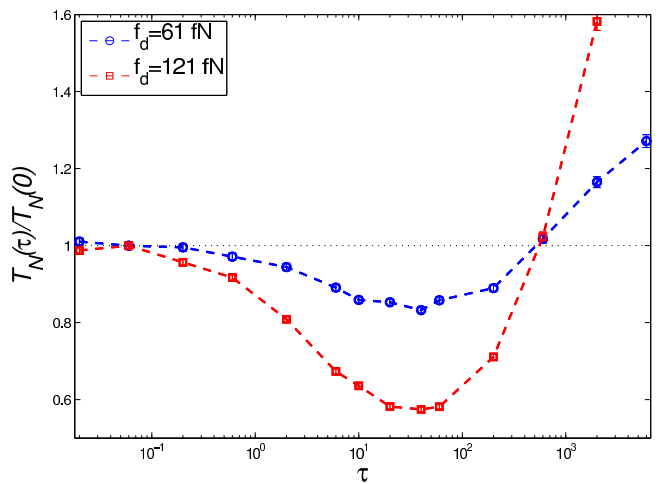}
\caption{The normalized looping time $\mathcal{T}_{N}(\tau)/ \mathcal{T}_{N}$ in the presence of dichotomic tension for the chain of $N=12$. The dichotomic force amplitudes are $f_d=61$ (circle) and $f_d=121$  fN (rectangular). The looping time shows a minimum at optimal flipping time $\tau_{r}\approx40$. \label{fig:N12_T}}
\end{figure}

To study how the dichotomic force affects the RA phenomena, we consider different value of $f_d$. Figure~\ref{fig:N12_T} shows the normalized looping time with $f_{d}=61$ (circles), 121 fN (squares) \textit{in the absence} of static tension ($f=0$). The looping time has resonant minimum at the optimum flipping time $\tau_{r}\approx40$ which is smaller than the case with static tension $f$. At the optimum flipping time, the looping time $\mathcal{T}_{N}(0,\tau)$ is smaller for larger $f_d$, while, at very long $\tau$, $\mathcal{T}_{N}(0,\tau)$ is larger for larger $f_d$. Therefore at certain $\tau$ in between, there is a crossing point.

\begin{figure}
\includegraphics[width=8.5cm] {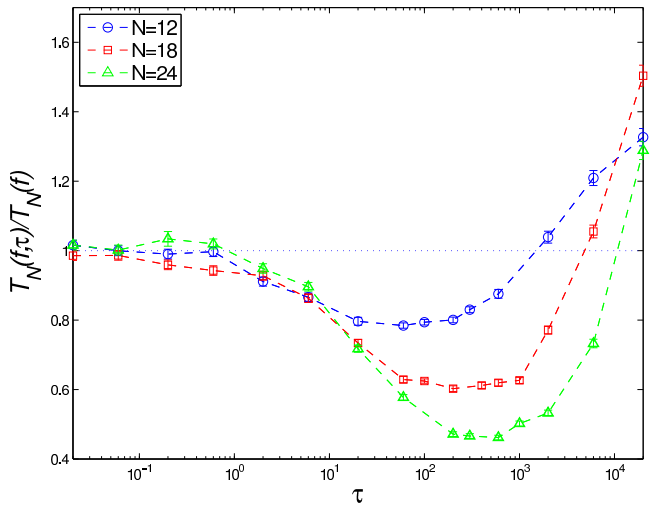}
\caption{The normalized looping time $\mathcal{T}_{N}(f, \tau)/\mathcal{T}_{N}(f)$ in the presence of dichotomic tension as well as static tensions for the chains of $N=12$ (circle), 18 (rectangular), and 24 (triangle). The dichotomic force amplitude is $f_d=61$ fN, and the static tension amplitude is $f=81$ fN. As $N$ increases, the optimal flipping times $\tau_{r}$ increase and the minimum values of normalized looping time $\mathcal{T}_{N}(f, \tau_{r})/\mathcal{T}_{N}(f)$ decrease.\label{fig:compare}}
\end{figure}

In contrast to the RA of single particle, the fluctuating barrier heights and thus the looping time depends sensitively on the chain length $N$. We obtain the normalized looping time $\mathcal{T}_{N}(f,\tau)/\mathcal{T}_{N}(f)$ for different chain lengths $L$ in the presence of dichotomic force of amplitude $f_{d}=61$ fN and static tension $f=81$ fN (Fig.~\ref{fig:compare}). While the optimal flipping time $\tau_r$ increases with $L$ as implied by the relation $\tau_{r}=\alpha L^{2}/D$, the minimum of $\mathcal{T}_{N}(f, \tau)/\mathcal{T}_{N}(f)$ decreases with $L$. This is because, similar to the static tension case, the fluctuating barrier height will be a function of $\beta f_{d}L$, so that the longer chain tends to be more susceptible to the tension $f_d$, and have the lower relative looping time.

A recent experiment~\cite{chen2010b} has shown that a fluctuating tension on DNA greatly reduces the looping time for small $\tau$, in consistency with our results. They attributed the phenomenon to an increase of the effective temperature. This may be reasonable for the small $\tau$, where the dichotomic noise adds to the thermal noise, but could not lead to the nonmonotonic resonant behavior emerging over the entire range of $\tau$.

\section{Conclusions}
\label{sec:conclusion}

We have studied the effects of static or time-dependent tension on the semiflexible polymer looping using Brownian dynamics simulation. For the case of static tension, we have found that a minute tension as small as 100 fN can dramatically change the looping time, especially for long chains. This sensitivity is a consequence of the cooperativity of the chain arising from chain connectivity. For the case of time-dependent tension, we considered dichotomically fluctuating tension, where the tension in average changes its sign in time $\tau$. The looping time has a resonant minimum at an optimum flipping time $\tau_r$, which is a nontrivial extension of the resonant activation (RA) of single Brownian particle. The effect of time-dependent tension is also more significant for longer chains. Our results are consistent with recent experiments for both static~\cite{chen2010a} and a fluctuating tension~\cite{chen2010b} cases. Although we neglect the details of chain conformation outside of the loop and orientation constraints, etc., our model could be a basic step to understand the loop formation process $\textit{in vivo}$ where biopolymers are constantly subject to forces. 
This study suggests a possibility that a biopolymer can self-organize optimally utilizing the ambient nonequilibrium fluctuations. Further experiment using dichotomically fluctuating tension with various $\tau$ is called for to establish the RA phenomena in polymer loop formation. 

\begin{acknowledgments}
This work was supported by Korea Research Foundation administered by Ministry of Education, Science and Technology, S. Korea.
\end{acknowledgments}


\end{document}